\pdfoutput=1
\documentclass[aip,jcp,citeautoscript,twocolumn,superscriptaddress,10pt]{revtex4-1}%

\usepackage{hyperref}

\usepackage{graphicx}
\usepackage[per-mode = symbol]{siunitx}
\usepackage{gensymb}
\usepackage{xspace}
\usepackage[version=4]{mhchem}
\usepackage[T1]{fontenc}
\usepackage{todonotes}

\newcommand\cst{\ensuremath{\langle\cos^2\theta_\text{2D}\rangle}\xspace}

\newcommand{\cfeldesy}{\affiliation{Center for Free-Electron Laser Science,
      Deutsches Elektronen-Synchrotron DESY, Notkestra{\ss}e 85, 22607 Hamburg,
      Germany}}%
\newcommand{\uhhcui}{\affiliation{The Hamburg Center for Ultrafast Imaging,
      Universit\"at Hamburg, Luruper Chaussee 149, 22761 Hamburg, Germany}}%
\newcommand{\uhhphys}{\affiliation{Department of Physics, Universit\"at Hamburg,
      Luruper Chaussee 149, 22761 Hamburg, Germany}}%

\begin{document}


\title{Switched Wave Packets with Spectrally Truncated Chirped Pulses}




\author{Adam S. Chatterley}
\affiliation{Department of Chemistry, Aarhus University, Langelandsgade 140, DK-8000 Aarhus C, Denmark}
\author{Evangelos T.\ Karamatskos}\cfeldesy\uhhphys%
\author{Constant Schouder}
\affiliation{Department of Chemistry, Aarhus University, Langelandsgade 140, DK-8000 Aarhus C, Denmark}
\author{Lars Christiansen}
\affiliation{Department of Chemistry, Aarhus University, Langelandsgade 140, DK-8000 Aarhus C, Denmark}
\author{Anders V. J\o rgensen}
\affiliation{Department of Chemistry, Aarhus University, Langelandsgade 140, DK-8000 Aarhus C, Denmark}
\author{Terry Mullins}\cfeldesy%
\author{Jochen K\"upper}\cfeldesy\uhhphys\uhhcui%
\author{Henrik Stapelfeldt} \email[]{henriks@chem.au.dk}
\affiliation{Department of Chemistry, Aarhus University, Langelandsgade 140, DK-8000 Aarhus C, Denmark}

\date{\today}

\begin{abstract}
A new technique for obtaining switched wave packets using spectrally truncated chirped laser pulses is demonstrated experimentally and numerically by one-dimensional alignment of both linear and asymmetric top molecules. Using a simple long-pass transmission filter, a pulse with a slow turn on and a rapid turn off is produced. The degree of alignment, characterized by \cst\ rises along with the pulse intensity and reaches a maximum at the peak of the pulse. After truncation \cst drops sharply but exhibits pronounced half and full revivals. The experimental alignment dynamics trace agrees very well with a numerically calculated trace based on solution of the time-dependent Schr\"odinger equation. However, the extended periods of field-free alignment of asymmetric tops following pulse truncation reported previously is not reproduced in our work.

\end{abstract}


\maketitle
\section{Introduction\label{intro}}

Quantum control of molecular dynamics is one of the ultimate goals of chemical
physics. Essentially, the aim is to create a custom wave packet with precisely
controlled population and phases of each eigenstate. To date, the most
successful approach for creating controlled wave packets is to use shaped
ultrafast laser pulses.~\cite{rabitz_whither_2000, brif_control_2010,
   brixner_quantum_2003} The construction of wave packets composed of states
ranging from rotational~\cite{bisgaard_observation_2004, renard_controlling_2004,
   renard_control_2005, horn_adaptive_2006,suzuki_optimal_2008} to
electronic~\cite{stolow_applications_1998, verlet_controlling_2003,
   calegari_ultrafast_2014}, has been demonstrated.

A special case of shaped laser pulses are truncated pulses which are switched
off much faster than they are switched on. The slow turn-on of the laser pulse
adiabatically transforms the field-free states of the molecule into the
eigenstates of the molecule in a dressed potential. The rapid truncation
nonadiabatically projects the field-dressed states onto their field-free
counterparts with well-defined phases. The archetypal example of this scheme,
termed switched wave packets, is in molecular alignment dynamics where the wave
packet is composed of rotational states populated by a non-resonant laser pulse.
They were theoretically introduced by Seideman~\cite{yan_photomanipulation_1999,
   seideman_dynamics_2001}, and realized experimentally by Stolow and
coworkers~\cite{underwood_switched_2003, sussman_quantum_2006}. Switched
rotational wave packets are especially interesting as (for symmetric tops) the
peak alignment acquired during the long pulse should reconstruct exactly during
the phase revivals, giving a route to extremely confined field-free alignment.
In effect, switched wave packets offer a best-of-both-worlds compromise between
adiabatic alignment, which enables a high degree of alignment at the cost of a
field present, and impulsive alignment, which enables field-free alignment, but
often with a lower degree of alignment~\cite{stapelfeldt_colloquium:_2003,
   seideman_nonadiabatic_2005}.

The generation of the necessary pulses with a very long turn-on and a very rapid
turn-off was, traditionally, quite difficult. The slow turn-on precluded the use
of standard liquid crystal or acousto-optic pulse shaping technology, and
instead a complex plasma shutter methodology was used~\cite{alcock_fast_1975},
involving nonlinear effects to truncate a long pulse with a synchronized
femtosecond pulse. Nevertheless, despite the interest in switched wave packet
alignment dynamics and the promises it offers, very few experimental
demonstrations were performed~\cite{underwood_switched_2003,
   sussman_quantum_2006, goban_laser-field-free_2008, mun_laser-field-free_2014,
   takei_laser-field-free_2016}. Recently, using chirped alignment
pulses~\cite{trippel_strongly_2013}, switched wave packet dynamics have been
demonstrated using non-Gaussian picosecond pulses with
short~\cite{trippel_strongly_2014} or long~\cite{trippel_two-state_2015,
   shepperson_observation_2018} turn-offs, but not with the combination of a
slow turn-on and a fast turn-off.

Here, we show a new, simple method for forming high-contrast truncated pulses
with only a single passive optic. We demonstrate switched wave packets in the
alignment dynamics of both linear molecules and asymmetric tops. The
experimental dynamics match very well with numerical solutions to the
time-dependent Schr\"odinger equation, demonstrating that the pulse is well
characterized and the dynamics can be completely explained by the interference
of rotational eigenstates.

\section{Methods\label{methods}}
\subsection{Experiment}

The experimental setup is identical to our previous chirped pulse adiabatic alignment experiments~\cite{shepperson_strongly_2017,shepperson_observation_2018}, with the exception of the addition of a longpass interference filter into the alignment beampath. Carbonyl sulfide (OCS) or iodobenzene (IB) is seeded in 80 bar helium and expanded through an Even-Lavie pulsed valve, giving an estimated rotational temperature of \SI{1}{\kelvin}~\cite{even_cooling_2000,nielsen_stark-selected_2011,shepperson_observation_2018}. The uncompressed output of a Ti:sapphire chirped pulse amplifier (\SI{160}{\pico\second} FWHM, GDD = +\SI{1.3}{\pico\second^2}) is truncated using a longpass filter to form the spectrally truncated chirped pulse (STCP, see below), which aligns molecules and prepares a switched wave packet. The compressed portion of the beam (\SI{35}{\femto\second} FWHM) is focused to an intensity of \SI{4E14}{\watt/\centi\metre^2} and probes the molecules by Coulomb explosion. The degree of alignment of the molecules is measured by recording the expectation value \cst of the projection of the emission angle of \ce{S+} or \ce{I+} ions relative to the alignment polarization using velocity map imaging. The degree of alignment is quantified by computing the expectation value \cst.The STCP was linearly polarized parallel to the detector plane, and the femtosecond probe pulse was polarized along the detector normal. The probe focal spot size was significantly smaller ($\omega_\text{0}$ = \SI{20}{\micro\metre}) than the STCP focal spot size ($\omega_\text{0}$ = \SI{38}{\micro\metre}) to minimize focal volume effects.

\subsection{Theory}

To characterize the experimental results, the degree of alignment was numerically simulated by solving the time-dependent Schr{\"o}dinger equation using the experimental alignment-pulse profiles. Two separate codes were used to calculate the rotational dynamics for the linear-top molecule OCS~\cite{sondergaard_nonadiabatic_2017} and the asymmetric-top molecule iodobenzene~\cite{omiste_rotational_2011, thesing_time-dependent_2017}, respectively. In brief, the rotational part of the Schr{\"o}dinger equation was solved for the rigid-rotor coupled to a non-resonant laser field, where relativistic, fine- and hyperfine interactions as well as nuclear spin effects were neglected. For iodobenzene, the asymmetric top wave functions were expanded in a symmetry-adapted basis built of symmetric rotor states~\cite{thesing_time-dependent_2017}. In both cases, the simulations assumed a rotational temperature of 1 K and included focal volume averaging determined by the experimental spot sizes of the laser beams. The 2D projection of the degree of alignment, \cst, was computed to provide a direct comparison with them experimental results.

\section{Results and Discussion\label{results}}
\subsection{Spectrally Truncated Chirped Pulses\label{STCP}}

Key to switched wave packet methodologies are laser pulses with a slow turn-on and a rapid turn-off, where `slow' and `rapid' are taken in comparison to the characteristic timescale of the wave packet in question. For plasma shutters, the turn-on time is defined by the pulse duration of the long pulse, and the turn-off time is defined by the duration of the short pulse, allowing nanosecond turn-ons, with tens of femtosecond turn-offs. However, plasma shutters are not without limitations. The required optical setup is fairly elaborate and sensitive, as it depends on highly nonlinear interactions. Additionally, without great care there is a residual field after truncation, which  in most demonstrated applications has had an intensity of several percent of the peak before truncation~\cite{underwood_switched_2003,goban_laser-field-free_2008}. This significant residual field means that the truncation does not project wave packets onto true field-free eigenstates, but rather modestly coupled ones.

By working with highly chirped pulses in the spectral domain, truncated pulses
can be produced that overcome the limitations of plasma shutters, while also
allowing for a much simpler optical setup. Highly chirped broadband pulses have
previously been used for adiabatic and intermediate regime
alignment~\cite{trippel_strongly_2013, trippel_strongly_2014,
   kierspel_strongly_2015, trippel_two-state_2015,
   chatterley_three-dimensional_2017, Trippel:JCP:inprep}, and our truncation
technique is a natural extension. Its essence comes in the observation that for
broadband laser pulses with large amounts of chirp (second order dispersion),
the time-domain pulse shape has a 1:1 correspondence with the spectrum. This can
be intuitively verified by considering the definition of second order
dispersion, which is the time delay between adjacent frequency components. This
implies that the temporal profile of the chirped pulses can be directly shaped
by modifying the spectrum with a device as simple as an interference
transmission filter. Truncation of a positively (negatively) chirped pulse can
be achieved through simple application of a long- (short-) pass filter.

\begin{figure}
\includegraphics[width=8.5 cm]{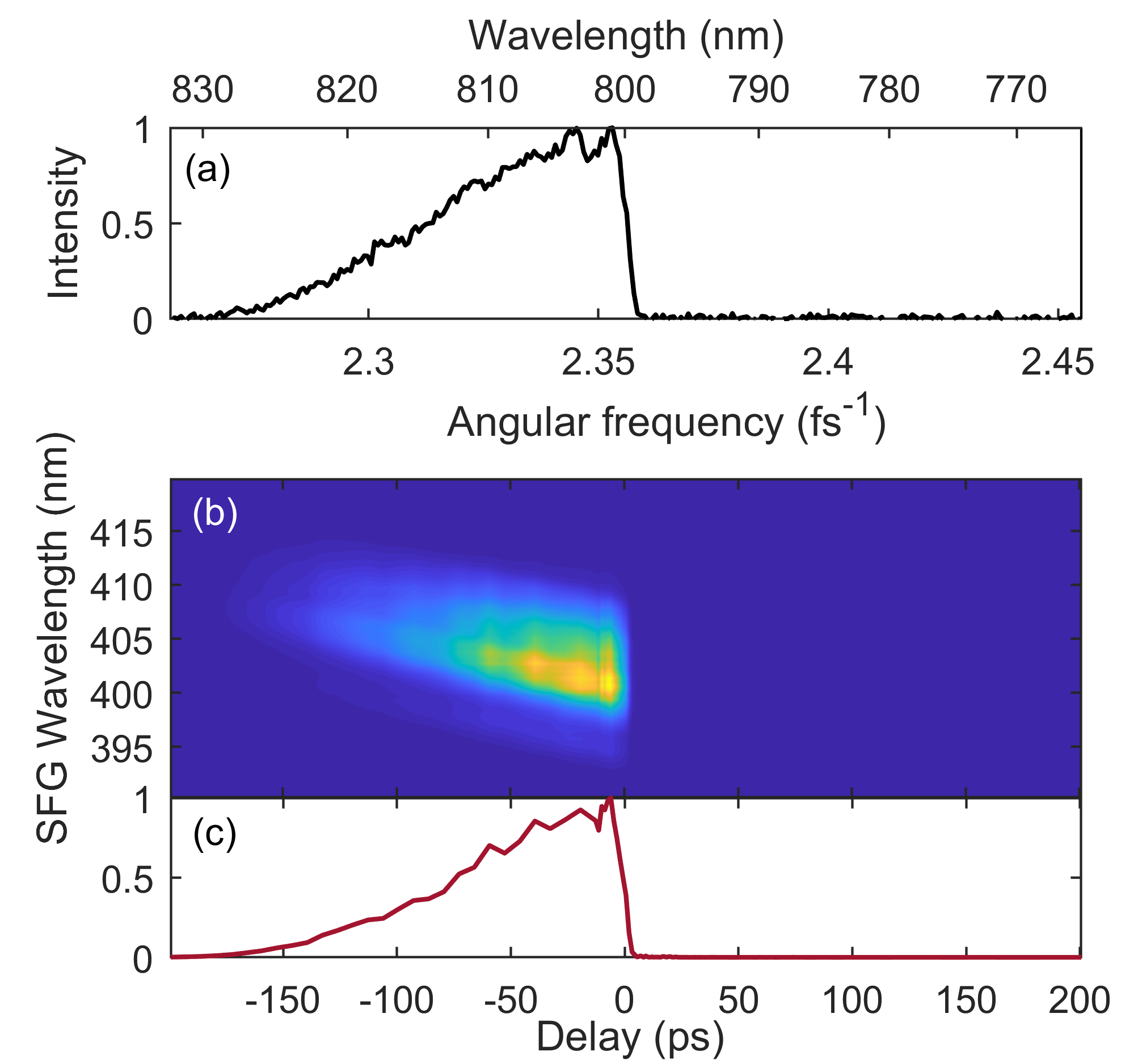}%
\caption{\label{pulse} (a) Spectrum of the uncompressed output of a Ti:sapphire chirped pulse amplifier after transmission through an 800 nm long-pass filter. (b) Spectrally resolved sum-frequency cross correlation of the truncated chirped pulse with a short 800 nm pulse. (c) Temporal profile of the intensity of the spectrally truncated chirped pulse. Time zero has been defined as half of maximum intensity on the falling edge. }
\end{figure}

Fig.~\ref{pulse}(a) shows the spectrum of the uncompressed (positively chirped) output of a Ti:sapphire chirped pulse amplifier, after transmission through an \SI{800}{nm} longpass spectral filter (Thorlabs FELH0800). The spectrum is abruptly cut at 800 nm. The influence of spectrum on temporal profile is shown in Fig.~\ref{pulse}(b), which shows the frequency-resolved sum frequency generation cross correlation between the truncated chirped pulse and a transform limited \SI{35}{\femto\second}, \SI{800}{\nano\metre} short pulse. The arrival time of each wavelength increases linearly with frequency, until the 800 nm component is reached, when the pulse is abruptly truncated. The temporal profile, found by integrating Fig.~\ref{pulse}(b) over the wavelength, is shown in Fig.~\ref{pulse}(c). The turn-on time for this pulse is around \SI{150}{\pico\second} (10--90~\%), while the turn-off time (90--10~\%) is approximately \SI{8}{\pico\second}. The residual intensity is  2--3 orders of magnitude lower than the pulse peak immediately after truncation, and undetectable after \SI{30}{ps} (see supporting information). Compared to pulses truncated using a plasma shutter, spectrally truncated chirped pulses (STCPs) offer high post-truncation contrast and a simple generation method. The drawback is that the truncation time is limited by the sharpness of the spectral filter, which in this case limits us to an \SI{8}{\pico\second} turn-off time. As we show below, this turn-off time is still sufficiently fast to produce switched wave packet dynamics, even in light rotors such as OCS. If sharper edges are required, they could be obtained by replacing the transmission filter with a pulse shaper based on diffractive optics and spatial filtering, i.e., diffraction gratings and a razor blade~\cite{weiner_ultrafast_2011}. Such a setup would be more complex than the simple transmission filter, but could likely reduce the turn-off time to a few ps or less.

\subsection{Alignment of Linear Molecules \label{OCS}}

\begin{figure}
\includegraphics[width=8.5 cm]{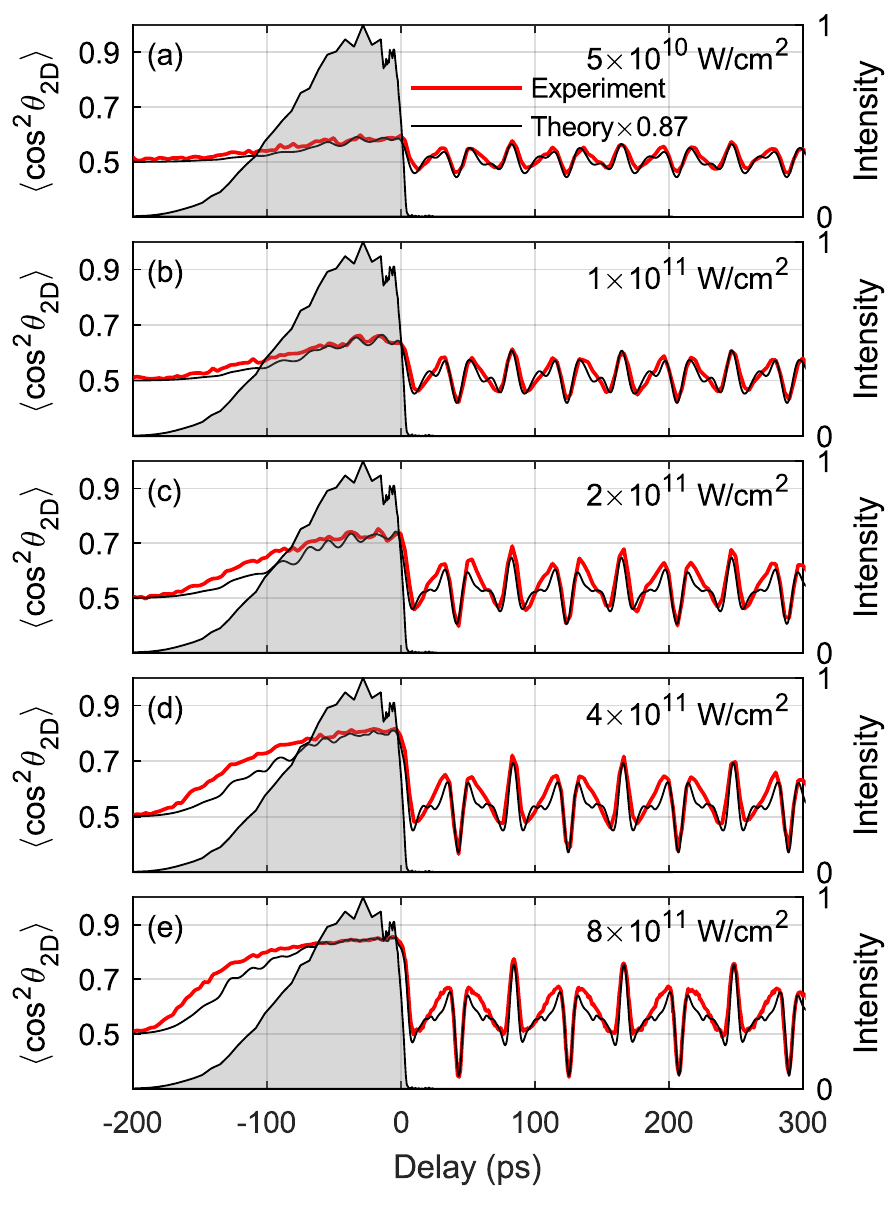}%
\caption{\label{OCSFIG} Alignment dynamics of OCS with a spectrally truncated chirped pulse (shaded area) with increasing peak intensity, found experimentally (thick red) and numerically (thin black). The experimental \cst is determined using the angular distribution of \ce{S+} fragments following Coulomb explosion. The peak alignment intensity is given in the top right of each plot. To account for non-axial recoil in the experimental \cst, the numerical results have been scaled around 0.5 by an arbitrary value of 0.87. }
\end{figure}

Our first demonstration of alignment using STCPs is with the linear OCS molecule. Alignment (and orientation) dynamics of OCS have previously been well studied, with both impulsive alignment~\cite{loriot_laser-induced_2007,nielsen_stark-selected_2011,kienitz_adiabatic_2016,fraia_impulsive_2017},  switched wave packet methodologies~\cite{goban_laser-field-free_2008}, and intermediate regimes with moderately long alignment pulses~\cite{trippel_strongly_2014,trippel_two-state_2015}, so a direct comparison is possible. Fig.~\ref{OCSFIG} shows as red curves the value of \cst as a function of delay relative to the STCP (shaded area), for peak alignment intensities ranging from $\SI{5e10}{W/cm^2}$ to $\SI{8e11}{W/cm^2}$. The value of \cst is determined by observing the recoil direction of \ce{S+} fragments following Coulomb explosion with the probe pulse. As expected, the degree of alignment increases smoothly during the pulse turn-on, reaching a maximum of \cst = 0.85 for the $\SI{8e11}{W/cm^2}$ pulse (Fig.~\ref{OCSFIG}(e)). The peak of alignment is similar to that observed previously with \SI{50}{ps} and \SI{500}{ps} pulsed and a similar intensity~\cite{trippel_strongly_2014,trippel_two-state_2015}. Following truncation, a clear oscillating revival structure is observed, with a full period of \SI{82.2}{ps}. As the alignment intensity is increased, the revivals become sharper, and the average alignment post-truncation increases.

To explain the observed alignment dynamics, we turn to numerical simulations
(black curves in Fig.~\ref{OCSFIG}). In Fig.~\ref{OCSFIG}, the displayed values of \cst of the simulated data
have been scaled by an arbitrary factor of 0.87, symmetrically centered around 0.5, as a simple means to account for non-axial recoil in the Coulomb explosion process depressing the experimental values~\cite{christensen_deconvoluting_2016}. Except for a slight discrepancy on the rising edge of the pulse, the numerical simulations capture the experimental dynamics remarkably well. The shape of the revivals are well reproduced, and there is even a capturing of the slight oscillatory structure visible on the rising edges of the lower intensity curves. The quality of agreement between theory and experiment gives us a high degree of confidence both of the accuracy of the alignment pulse measurement, and of our ability to simulate alignment dynamics with these complex pulses.

The change of the shape of the alignment revivals as intensity increases is a direct reflection of the number of rotational $J$ states populated during the pulse turn-on. The lowest intensity trace (Fig.~\ref{OCSFIG}(a)) has an almost sinusoidal pattern, suggesting a few-state wave packet similar to that previously observed when exciting with a low intensity, \SI{50}{ps} long pulse~\cite{trippel_strongly_2014}. On the other hand, the most intense trace, Fig.~ \ref{OCSFIG}(e), has a much more complex structure, suggesting rather more frequency components. This is born out in the numerical simulations: from the ground rotational state, the lowest intensity trace reaches a maximum angular momentum $J_{max} = 4$, while the highest has $J_{max} = 8$, twice as many rotational states contribute. Note, however, that these values of  $J_{max}$ are very low compared to typical impulsive alignment experiments, where a simple simulation finds $J_{max} > 30$ when kicked with a typical \SI{10}{TW/cm^2}, \SI{500}{fs} pulse~\cite{sondergaard_nonadiabatic_2017}. This highlights the power of switched wave packets to reach a previously unexplored regime of alignment dynamics, where very high degrees of alignment are obtained with few populated rotational states, presumably because of a much stronger phase relationship than is found in impulsive alignment, where many more rotational states are populated~\cite{stapelfeldt_colloquium:_2003}. The phases of wave packets produced by STCPs will be examined in detail in a forthcoming publication.

In the limit of instantaneous pulse truncation, we would expect that at each full revival the wave packet should reconstruct exactly, and \cst should be identical to the peak of alignment prior to truncation~\cite{yan_photomanipulation_1999,seideman_dynamics_2001,underwood_switched_2003}. In our case the peak of alignment during the full revivals reaches $\sim90\%$ of the peak alignment, for the highest intensity, and $\sim95\%$ for the lowest intensity, indicating that the pulse truncation is not a completely impulsive process, and that some reduction in angular momentum must occur during the truncation. This result is unsurprising, as the \SI{8}{\pico\second} truncation time is not negligible when compared to the \SI{82.2}{\pico\second} rotational period of OCS~\cite{trippel_strongly_2014}. The fact that lower intensity pulses lead to better reconstruction suggests that it is the higher $J$ states which are mostly depopulated by the non-instantaneous turn-off of the alignment pulse. We note, however, that switched wave packet experiments using rapid truncation with plasma shutters also have not achieved complete wave packet reconstructions at the peak of revivals; this is likely an effect of the residual field following truncation with that method~\cite{underwood_switched_2003, goban_laser-field-free_2008}.

\subsection{Alignment of Asymmetric Tops\label{IB}}

Although alignment of linear molecules provides a simple test case for STCP dynamics, in reality most molecules are asymmetric tops with much more complex rotational dynamics. Asymmetric tops have three rotational constants which contribute to alignment dynamics, and in general their revival structures are much richer than linear or symmetric top molecules. To benchmark the behaviour of asymmetric tops when aligned with STCPs, we have measured and calculated the alignment dynamics of iodobenzene (IB). IB has been extensively studied using both impulsive~\cite{peronne_nonadiabatic_2003, poulsen_nonadiabatic_2004, bisgaard_observation_2004, holmegaard_control_2007, ren_measurement_2012} and adiabatic alignment~\cite{poulsen_photodissociation_2002,trippel_strongly_2013}, and was recently explored using a plasma shutter truncated pulse by Sakai and co-workers, who observed a very surprising persistent alignment for many picoseconds following truncation~\cite{mun_laser-field-free_2014}.

\begin{figure}
\includegraphics[width=8.5 cm]{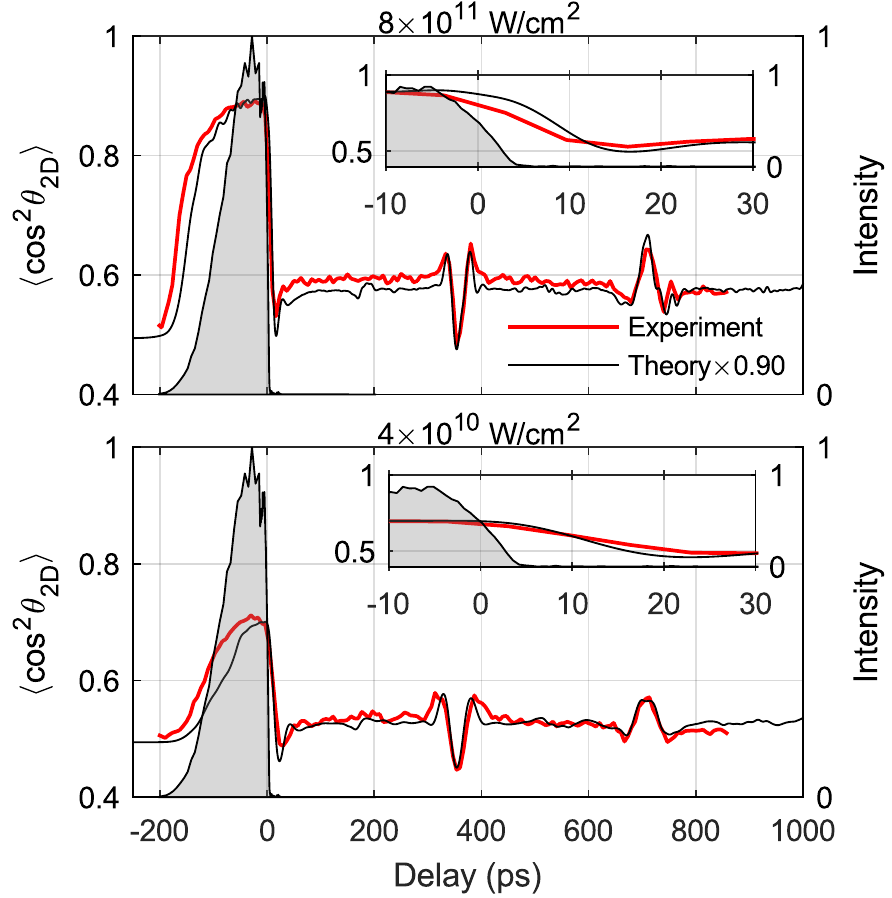}%
\caption{\label{IBFIG} Alignment dynamics of iodobenzene with truncated pulses. The pulse intensity profile is given by the shaded area. Experimental values of \cst are shown as thick red lines, while the calculated dynamics are given by the thin black lines, for (a) $\SI{8e11}{W/cm^2}$ and (b) $\SI{4e10}{W/cm^2}$ peak alignment intensities. To account for non-axial recoil in the experiment, calculated curves have been scaled by 0.90, symmetrically around 0.5.}
\end{figure}

Fig.~\ref{IBFIG} shows (as thick red lines) the alignment dynamics of IB, when subjected to STCPs of either $\SI{8e11}{W/cm^2}$ (a) or $\SI{4e10}{W/cm^2}$ (b). The inset expands the region around the truncation point. Simulations, using the experimental pulse, are overlaid as thin black lines. Similar to the OCS simulations, these have been scaled down to account for non-axial recoil in the experimental data, in this case by a factor of 0.9.  Following truncation, the degree of alignment abruptly decreases, then returns to a permanent alignment of $\cst\sim~0.59$ and 0.53 for the high and low intensity STCP, respectively. Prominent, broad revival features are observed, centered at 359 ps and 712 ps. The agreement between the experiment and theory is exceptional, all major features in the experimental trace are accurately reproduced by the theory.

We first consider the behaviour of the revivals in the field-free regime. Compared to IB dynamics following impulsive alignment~\cite{holmegaard_control_2007}, several striking aspects are immediately observable. Firstly, only $J$-type revivals are observed, which correspond to symmetric-top like motion. With high intensity impulsive alignment, $C$-type revivals, which correspond to motion around the axis perpendicular to both the molecular plane and the C-I bond are visible at 380 and 757 ps, however we see no evidence for these  with either high or low alignment intensity STCPs. $C$-type revivals are associated with high lying rotational states~\cite{rouzee_field-free_2006}, so it is reasonable that they are not observed with our adiabatically prepared alignment. Secondly, the directions of the revivals are inverted: with impulsive alignment the half-revival is a positive feature, and the full revival negative, the exact inverse of the dynamics we observe. This inversion presumably reflects the different preparation methods and the precise phases of the components in the wave packets; this effect will be investigated in greater detail in a forthcoming publication. Finally, the STCP revivals are exceedingly broad, compared to impulsive revivals. The main peaks of both the full- and half-revivals are around 50 ps wide, compared to around 20 ps when excited with a \SI{200}{fs}, \SI{16}{TW/cm^2} kick~\cite{holmegaard_control_2007}. The breadth of these revivals is further evidence that very few rotational states are populated when truncated pulses are used for alignment - there simply are not enough Fourier components to construct a sharp feature.

\begin{figure}
\includegraphics[width=8.5 cm]{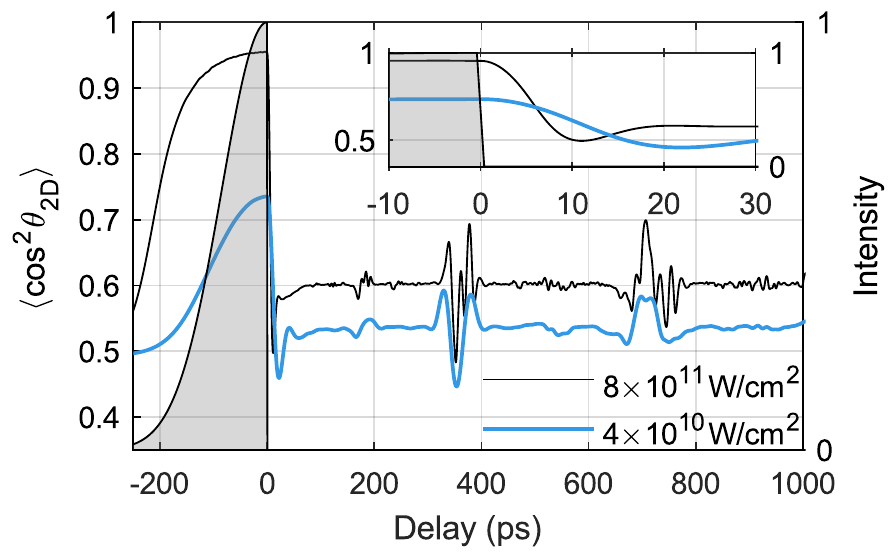}%
\caption{\label{IBTHEORY} Calculated values of \cst with a \SI{<200}{fs} rapidly truncated alignment pulse with peak intensity of $\SI{8e11}{W/cm^2}$ (thin black line), $\SI{4e10}{W/cm^2}$ (thick blue line).  }
\end{figure}

We now turn to the behavior of the alignment at the moment of pulse truncation (inset Fig.~\ref{IBFIG}). As soon as the pulse truncation is complete, ($t$ = 5 ps), the degree of alignment drops appreciably. With a high intensity STCP, the alignment reaches a minimum at  $t = \SI{18}{ps}$, while low intensity STCPs give a minimum at $t = \SI{28}{ps}$. The theory captures the general trend, although somewhat overestimates the speed of the alignment drop for the high intensity, and underestimates it for the low intensity. This discrepancy is likely due to either temperature or focal volume effects differing slightly between simulation and experiment. The slower drop in alignment with low intensity can again be explained by the lower number of populated rotational states giving insufficient high frequency Fourier components for a fast drop. This behavior, however, stands in stark contrast to the dynamics observed by Sakai and coworkers, using a pulse shaped by a plasma shutter~\cite{mun_laser-field-free_2014}. They reported a high degree of alignment that appeared to persist for $\sim\SI{10}{ps}$ after the pulse was very abruptly ($<\SI{200}{fs}$) truncated. Could it be the case that very abrupt pulse truncation leads to significant field-free alignment, which is not present if the pulse turns off too slowly? To investigate this possibility we have also performed simulations, with identical conditions and alignment intensities to our experiment, but with the turn-off time of the pulse reduced to \SI{150}{\femto\second}, shown in Fig.~\ref{IBTHEORY}. The excellent experimental agreement of our simulations with a slow turn-off time, Fig.~\ref{IBFIG}, give us confidence that our calculations accurately and comprehensively model the alignment dynamics of asymmetric top systems, and so we fully trust our simulations with rapidly truncated pulses. On the whole, the alignment dynamics are similar to those with the longer turn-off time, however the revivals are more structured and show evidence of $C$-type behavior, particularly with high alignment intensity. This demonstrates that the rotational states populated during the adiabatic process are better preserved during rapid truncation, in agreement with our findings from OCS and the predictions of Seideman~\cite{seideman_dynamics_2001}.

Compared to the experimental STCP, the falling edge of the alignment is faster, in agreement with the intuitive expectation that the quicker the field is removed, the sooner molecules should no longer be aligned by it. The alignment begins to drop appreciably within 2 ps of the field truncation, regardless of whether a high or a low intensity alignment field is used. The slope is less severe with a weak alignment intensity, but as this necessitates a weak degree of alignment, this effect cannot be exploited for high quality field-free alignment. These findings lead us to suspect that further experimental aspects, such as possibly the residual post-truncation field, may have led to the reported findings by Sakai and co-workers~\cite{mun_laser-field-free_2014}. That group also reported an absence of revivals in their IB alignment dynamics~\cite{mun_laser-field-free_2015}, which are observed very clearly by us in both theory and experiment. It is feasible that whatever experimental complication led to a lack of observable revivals also led to apparent long-lived field-free alignment after truncation.

\section{Conclusions\label{conclusion}}

We have developed a new scheme for creating truncated laser pulses for preparation of switched wave packets, by spectrally filtering chirped laser pulses. Compared to the previous standard methodology for producing truncated pulses, our method has a trivial optical setup, high repetition rates, very good contrast after the pulse is truncated, and is directly applicable to free-electron laser experiments~\cite{kierspel_strongly_2015}. The primary drawback is a slower truncation time, which could be significantly enhanced if a more complex spectral filter was employed.

Using the newly developed STCPs to align the linear OCS molecule, we have observed high degrees of alignment while the alignment pulse is present, and rich rotational dynamics after the pulse is truncated. Compared to traditional impulsive alignment, very few rotational states are populated, leading to simple alignment dynamics. Numerical simulations, using the experimentally determined pulse shape, accurately reproduce the alignment dynamics. The degree of alignment at the revival is close to that during the maximum of adiabatic alignment, and it could be enhanced further by, for example, a scheme to `kick' the molecule at the peak of a revival~\cite{guerin_ultimate_2008}.

In iodobenzene, an asymmetric top system, we again observe excellent alignment during the pulse, and after truncation we see very broad $J$-type revivals, composed of few rotational levels. Computations accurately predicts these, even with the complex pulse shape involved. The alignment begins to decay as soon as the pulse is truncated, in contrast to previously reported experiments with IB and truncated pulses where field-free alignment was observed. By comparison with simulation, we suspect that those findings may have been due to an experimental artefact.

Our results pave the way for new investigations on switched wave packets. Since the invention of the methodology over a decade ago, switched wave packets have been of great interest to the community, but rather few experiments have been performed. With the development of the trivial STCP methodology, many laboratories will be able to investigate switched wave packets. Additionally, STCPs are extremely useful for molecular alignment, even without exploiting switched wave packets. If a superfluid helium solvent is present, then alignment persists for a brief time after the pulse is truncated, giving a simple route to field-free molecular alignment for several picoseconds~\cite{chatterley_three-dimensional_2017}. We have very recently employed helium droplets and STCPs to perform field-free 3D alignment on a range of complex molecules.

\section{Acknowledgements\label{acks}}

We acknowledge support from the European Research Council-AdG (Project No. 320459, DropletControl). This research was undertaken as part of the ASPIRE Innovative Training Network, which has received funding from the European Union's Horizon 2020 research and innovation programme under the Marie Sklodowska-Curie grant agreement No 674960. This work has been supported by the Deutsche Forschungsgemeinschaft (DFG) through the priority program 1840 ``Quantum Dynamics in Tailored Intense Fields" (QUTIF, KU1527/3) and the excellence cluster ``The Hamburg Center for Ultrafast Imaging -- Structure, Dynamics and Control of Matter at the Atomic Scale'' (CUI, EXC1074), and through the Helmholtz Association ``Initiative and Networking Fund''.


%

\end{document}